\documentclass[sigconf,natbib=true,anonymous=false]{acmart}

\AtBeginDocument{%
  }

\setcopyright{acmlicensed}
\copyrightyear{2026}
\acmYear{2026}
\acmDOI{XXXXXXX.XXXXXXX}
\acmConference[Acronym 'XX]{Make sure to enter the correct
conference title from your rights confirmation email}{June 03--05,
2018}{Woodstock, NY}

\acmISBN{978-1-4503-XXXX-X/2018/06}

\usepackage{amsmath} 
\usepackage{amsfonts} 
\usepackage{mathrsfs}
\usepackage{enumitem}
\usepackage{booktabs}  
\usepackage{tcolorbox}
\usepackage{multirow} 
\usepackage{graphicx}  
\usepackage{amsmath}  
\usepackage{mathrsfs}  
\usepackage{subcaption}
\usepackage{caption}
\usepackage{subcaption}

\usepackage{algorithm}
\usepackage{algorithmic}

\begin{document}

\title{RCLRec: Reverse Curriculum Learning for Modeling Sparse Conversions in Generative Recommendation}

\author{Yulei Huang}
\orcid{0009-0004-0378-5330}
\authornote{Contributed equally to this research.} 
\affiliation{
  \institution{Alibaba International Digital Commerce Group}
  \city{Hangzhou} 
  \state{} 
  \country{China}
}
\email{huangyulei.hyl@alibaba-inc.com}

\author{Hao Deng}
\orcid{0009-0002-6335-7405}
\authornotemark[1]
\affiliation{%
  \institution{Alibaba International Digital Commerce Group}
   \city{Beijing} 
   \state{} 
   \country{China}
}
\email{denghao.deng@alibaba-inc.com}

\author{Haibo Xing}
\orcid{0009-0006-5786-7627}
\affiliation{%
  \institution{Alibaba International Digital Commerce Group}
  \city{Hangzhou} 
  \state{} 
  \country{China}
}
\email{xinghaibo.xhb@alibaba-inc.com}

\author{Jinxin Hu}
\orcid{0000-0002-7252-5207}
\authornote{Corresponding author.}
\affiliation{
  \institution{Alibaba International Digital Commerce Group}
  \city{Beijing} 
  \state{} 
  \country{China}
}
\email{jinxin.hjx@lazada.com}

\author{Chuanfei Xu}
\orcid{0000-0003-2653-0387}
\affiliation{%
  \institution{AI, Guangdong Laboratory of Artificial Intelligence and Digital Economy}
  \city{Guangzhou} 
  \state{} 
  \country{China}
}
\email{xuchuanfei@gml.ac.cn}

\author{Zulong Chen}
\orcid{0000-0003-0807-6889}
\affiliation{%
  \institution{Alibaba Group}
  \city{Hangzhou} 
  \state{} 
  \country{China}
}
\email{zulong.czl@alibaba-inc.com}

\author{Yu Zhang}
\orcid{0000-0002-6057-7886}
\affiliation{
  \institution{Alibaba International Digital Commerce Group}
  \city{Beijing} 
  \state{} 
  \country{China}
}
\email{daoji@lazada.com}

\author{Xiaoyi Zeng}
\orcid{0000-0002-3742-4910}
\affiliation{
  \institution{Alibaba International Digital Commerce Group}
  \city{Hangzhou} 
  \state{} 
  \country{China}
}
\email{yuanhan@taobao.com}

\renewcommand{\shortauthors}{Trovato et al.}

\begin{abstract}
Conversion objectives in large-scale recommender systems are sparse, making them difficult to optimize. Generative recommendation (GR) partially alleviates data sparsity by organizing multi-type behaviors into a unified token sequence with shared representations, but conversion signals remain insufficiently modeled. While recent behavior-aware GR models encode behavior types and employ behavior-aware attention to highlight decision-related intermediate behaviors, they still rely on standard attention over the full history and provide no additional supervision for conversions, leaving conversion sparsity largely unresolved. To address these challenges, we propose \textbf{RCLRec}, a reverse curriculum learning–based GR framework for sparse conversion supervision. 
For each conversion target, RCLRec constructs a short curriculum by selecting a subsequence of conversion-related items from the history in reverse. Their semantic tokens are fed to the decoder as a prefix, together with the target conversion tokens, under a joint generation objective.
This design provides additional instance-specific intermediate supervision, alleviating conversion sparsity and focusing the model on the user's critical decision process. We further introduce a curriculum quality-aware loss to ensure that the selected curricula are informative for conversion prediction. 
Experiments on offline datasets and an online A/B test show that RCLRec achieves superior performance, with +2.09\% advertising revenue and +1.86\% orders in online deployment.

\end{abstract}




\maketitle
\begin{figure}[t]
  \captionsetup{skip=1pt}
  \centering
  \includegraphics[width=0.95\linewidth]{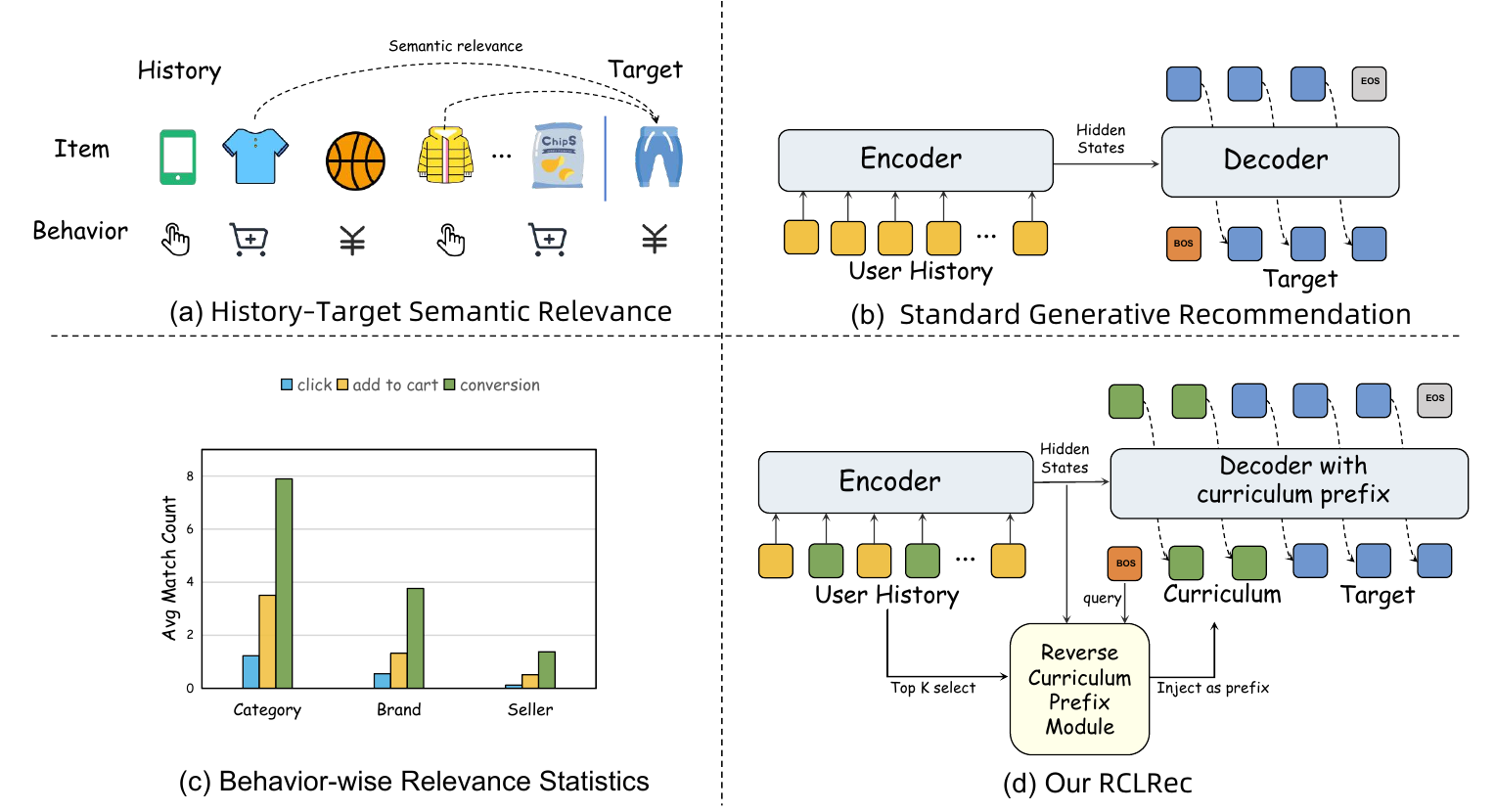} 
  \caption{(a) Multi-behavior history–target semantic relevance. (b) Standard GR method. (c) Behavior-wise relevance statistics showing that conversions are preceded by more coherent clusters of related items. (d) Our proposed RCLRec.}
\vspace{-18pt}
  \label{fig:intro}
\end{figure}
\section{Introduction}
Recently, generative recommendation (GR) based on semantic tokens has attracted extensive attention in both academia and industry \cite{reg4rec,hou2025survey,yang2024unifying}. Unlike traditional sequential recommenders \cite{sasrec, csmf}, GR organizes a user's multi-type behaviors (\textit{e.g.}, impressions, clicks, purchases) into a unified item sequence, and represents each item in this sequence using shared discrete semantic tokens obtained by vector quantization \cite{lee2022autoregressive, rpg, hou2023learning}  (see Figure~\ref{fig:intro}(b)). 
By sharing token-level representations across behavior types and exploiting large-scale interaction data, GR has achieved strong retrieval performance \cite{sid4rec,tiger,rpg}. However, behavior types in GR exhibit highly imbalanced distributions. While conversion behaviors (\textit{e.g.}, purchases) are much less frequent than impressions and clicks, they directly determine key business metrics such as revenue and orders \cite{csmf,zheng2022multi}. This \textbf{data sparsity} leaves the conversion objective insufficiently modeled by existing GR methods, and hinders accurate learning of user preferences at the conversion stage.

To better address this issue, recent work \cite{gear,mbgen,zhai2024actions, deng2025heterrec} explicitly incorporates behavior types into GR models. 
For example, some approaches encode behavior-type distinctions in the encoder and employ behavior-aware attention in the decoder \cite{mbgen,gear}. With this behavior-aware attention, GR can focus on highly indicative behaviors (\textit{e.g.}, revisits, comparisons, prior conversions) even under imbalanced behavior distributions. 
However, these methods do not fundamentally alleviate conversion sparsity, as the model still learns conversion patterns from very few events and receives \textbf{no additional supervision} specifically for the conversion objective.
Our empirical analysis shows that conversion events are typically preceded by a coherent cluster of related items (\textit{e.g.}, from the same category), reflecting a clearer and more focused decision process, as shown in Figure~\ref{fig:intro}(c). 
Thus, for conversion prediction, a small subset of behaviors in the long interaction history often forms a key decision-making subsequence that disproportionately influences the final purchase, providing a concentrated source of supervision for the sparse conversion objective.
Nevertheless, existing GR methods still rely on standard cross-attention to \textbf{implicitly discover useful intermediate behaviors} from the full history, without systematically identifying or explicitly modeling key subsequences for each conversion. As a result, these crucial signals remain buried among numerous irrelevant behaviors, so current GR methods struggle to fully exploit them when modeling conversions.

To address these problems, we propose \textbf{RCLRec} (Figure~\ref{fig:intro}(d)), a reverse curriculum learning–based GR framework specifically designed for sparse conversion supervision, drawing on curriculum strategies for sparse-reward reinforcement learning \cite{florensa2017reverse,xi2024training}. RCLRec uses reverse-constructed curricula of key intermediate behaviors to densify supervision signals. Unlike classical curriculum learning with multi-stage difficulty scheduling, we adopt reverse curriculum as a one-shot \textbf{decoder-side curriculum prefix}. This is because noisy, heterogeneous user histories \cite{schnabel2016recommendations,liu2018stamp} make it difficult to design reliable multi-stage curricula over interactions.
In detail, our main contributions are:
\begin{itemize}[noitemsep, topsep=0pt, leftmargin=*]
\item We propose a reverse curriculum prefix module (RCPM), which reversely selects 
$k$ key intermediate behaviors from the history
to form a curriculum prefix. Injected into the decoder alongside the target conversion tokens, this prefix provides additional instance-specific intermediate supervision and focuses the model on the user's critical decision process under sparse conversion signals.
\item We propose a curriculum quality-aware loss that encourages the selected subsequences to be informative for conversion.
\item We conduct experiments on two real-world datasets and an online A/B test, achieving significant improvements of +2.09\% advertising revenue and +1.86\% orders, demonstrating the practical effectiveness of RCLRec in production.
\end{itemize}

\section{Problem Formulation}

Let $\mathcal{U}$ and $\mathcal{I}$ denote the user and item sets. Let $\mathcal{B}$ denote the set of behavior types, 
such as \emph{click} (clk), \emph{add-to-cart} (atc), and \emph{conversion} (pay),
i.e., $\mathcal{B}=\{\text{clk},\text{atc},\dots,\text{pay}\}$.
For each user $u\in\mathcal{U}$, the interaction history is a chronological sequence
$S_u=\{(b_1,i_1),\dots,(b_T,i_T)\}$, where $b_t\in\mathcal{B}$, $i_t\in\mathcal{I}$, and $T$ is the sequence length.
Following recent GR methods~\cite{tiger,cobra}, we map each item $i\in\mathcal{I}$ to a length-$L$ token sequence $\mathbf{z}_i=[z_i^{1},\dots,z_i^{L}]$ using a discrete semantic mapping $\phi:\mathcal{I}\rightarrow \{V^\ell\}_{\ell=1}^L$, where $z_i^{\ell}\in\mathcal{V}^\ell$.
To support behavior-conditioned generation, we use a behavior embedding matrix $\mathbf{E}\in\mathbb{R}^{|\mathcal{B}|\times d}$ and treat $\mathbf{e}_b\in\mathbf{E}$ as the behavior-specific BOS control token in the decoder~\cite{mbgen} (\textit{e.g.}, $\mathbf{e}_{\text{pay}}$ for conversions).
Given $S_u$ and a target behavior $b$, GR retrieves items by generating the target token sequence $\mathbf{z}_{T+1}=\phi(i_{T+1})$ and minimizing the negative log-likelihood (NLL):
\begin{equation}
\mathcal{L}_{\text{GR}}
= -\sum_{\ell=1}^{L}\log p_\theta\!\left(z_{\text{T+1}}^{\ell}\mid S_u,\ \mathbf{e}_{b},\ \mathbf{z}_{\text{T+1}}^{<\ell}\right),
\label{eq:gr}
\end{equation}


where $\mathbf{z}_{\text{T+1}}^{<\ell}$ is the prefix of the target token sequence and $\theta$ are the model parameters.

\section{Methodology}

\subsection{Overview and Two-Stage Training}
\begin{figure*}[t]
  \centering
  \includegraphics[width=0.9\textwidth]{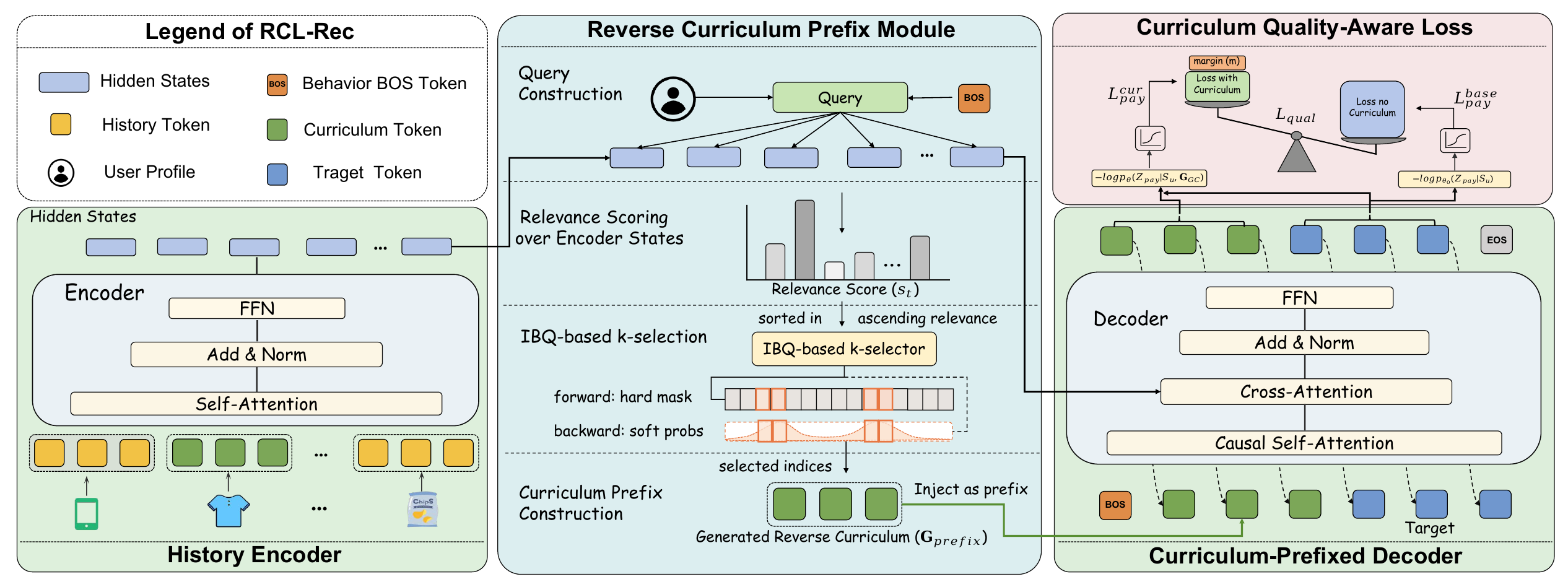}
  \caption{Overview of RCLRec. It uses a Reverse Curriculum Prefix Module (RCPM) to select conversion-relevant interactions from history as a decoder-side prefix, and is trained with a curriculum quality-aware loss.}
  \vspace{-12pt}
  \label{fig:method}
\end{figure*}

Figure~\ref{fig:method} illustrates the overall framework of RCLRec. To better learn sparse conversion targets, RCLRec introduces (i) \textbf{a reverse curriculum prefix module (RCPM)} that selects conversion-relevant historical interactions and injects their semantic tokens as a decoder prefix, and (ii) \textbf{a curriculum quality-aware loss function} that encourages the selected prefix to improve conversion likelihood (Sec. \ref{sec:loss}). We adopt a two-stage training scheme. Specifically, we first \textbf{pre-train} a standard encoder-decoder GR backbone on mixed multi-behavior data using Eq.~(\ref{eq:gr}) to learn general item semantic representations and behavior embeddings, including the conversion behavior token $\mathbf{e}_{\text{pay}}$. We then perform supervised fine-tuning (SFT) only on conversion samples, where we jointly optimize the SFT objective with RCPM and the curriculum quality-aware loss.

\subsection{Reverse Curriculum Prefix Module (RCPM)}
\label{sec:rcpm}
RCPM performs conversion-aware subsequence selection on the encoder side and exposes the selected items as decoder prefixes under teacher forcing. 
As shown in Figure~\ref{fig:intro}(c), conversions are often preceded by a short sequence of conversion-indicative behaviors in user interaction histories. 
Concretely, RCPM implements a reverse curriculum as a one-shot decoder-side prefix by tracing backward from the conversion target. 
Rather than simply selecting the most recent actions, it employs a pay-conditioned user query to retrieve those historical interactions that are most indicative of the conversion decision.
This reverse selection over heterogeneous user histories \cite{pan2026survey}, together with teacher forcing on the selected curriculum tokens, reduces interference from irrelevant behaviors and provides denser supervision and clearer context for conversion generation under sparse conversion signals.

Concretely, RCPM consists of two steps: (i) \textbf{reverse curriculum prefix selection} from the user’s interaction history, and (ii) an \textbf{end-to-end curriculum optimization mechanism}.

\subsubsection{Reverse curriculum prefix selection.}
\textbf{First}, to identify historical interactions that are most indicative of the conversion target, we construct a personalized query. Let $\mathbf{x}_u\in\mathbb{R}^{d_x}$ denote the feature vector of user $u$ (e.g., user ID, demographics), obtained via standard embedding and feature encoding \cite{csmf}. We map $\mathbf{x}_u$ to the model dimension as $\mathbf{e}_u=\mathbf{W}_u\mathbf{x}_u\in\mathbb{R}^{d}$, and reuse the conversion behavior embedding $\mathbf{e}_{\text{pay}}\in\mathbb{R}^{d}$ learned in pre-training. We concatenate them and feed the result into a lightweight multi-layer perceptron (MLP) to obtain the query:
$
\mathbf{q} = \text{MLP}\big([\mathbf{e}_u;\mathbf{e}_{\text{pay}}]\big)\in\mathbb{R}^{d}.
$

This user-specific, pay-conditioned query serves as the anchor for conversion-oriented curriculum selection. \textbf{Second}, given $\mathbf{q}$, we score the relevance of each historical interaction with respect to the conversion target using the contextual encoder states $\mathbf{H}_u=[\mathbf{h}_1,\ldots,\mathbf{h}_T]$. Each $\mathbf{h}_t$ encodes the $t$-th interaction and its contextual dependencies, reflecting its contribution to the user’s evolving intent. We compute the relevance score $s_t$ via a scaled inner product:
\begin{equation}
s_t = \frac{\mathbf{q}^\top \mathbf{h}_t}{\sqrt{d}},\quad t=1,\dots,T,
\end{equation}
where $d$ is the encoder hidden size. These scores $s_t$ are then normalized into a categorical distribution over the sequence:
\begin{equation}
p_t = \frac{\exp(s_t/\tau)}{\sum_{j=1}^{T}\exp(s_j/\tau)},\quad t=1,\dots,T,
\end{equation}
where $\tau>0$ is a temperature hyperparameter and $\mathbf{p}=[p_1,\dots,p_T]$ reflects the conversion-aware relevance of each historical behavior. \textbf{Third}, we select the top-$k$ behaviors for the reverse curriculum:
\begin{equation}
\mathcal{K} = \text{argmax}(\mathbf{p}, k) = \{t_1,\dots,t_k\},\quad |\mathcal{K}|=k,
\end{equation}
where $\text{argmax}(\cdot, k)$ returns the indices of the $k$ largest entries in $\mathbf{p}$. Each $t_j \in \mathcal{K}$ corresponds to a selected interaction $(b_{t_j}, i_{t_j})$, whose item tokens are used to construct the curriculum prefix. \textbf{Finally}, we construct the prefix by concatenating the semantic token sequences of the selected items $\phi(\mathcal{K})$ in ascending order of relevance scores:
\begin{small}
\begin{equation}
\mathbf{G}_{\text{prefix}} = \big[\phi(i_{t_1});\dots;\phi(i_{t_k})\big] 
= \big[\mathbf{z}_{i_{t_1}}^{1}, \dots,\mathbf{z}_{i_{t_1}}^{L},\dots,
\mathbf{z}_{i_{t_k}}^{1},\dots,\mathbf{z}_{i_{t_k}}^{L}\big],
\end{equation}
\end{small}
where the indices in $\mathcal{K}$ are sorted such that $p_{t_1} \le p_{t_2} \le \dots \le p_{t_k}$ and each $\phi(i_{t_j})$ is the semantic token sequence of item $i_{t_j}$. 
With this ordering, the most conversion-relevant item (with the largest $p_{t_k}$) appears last in the prefix and is therefore closest to the target tokens, providing stronger conditioning for conversion prediction.


During SFT, we inject $\mathbf{G}_{\text{prefix}}$ as a decoder-side prefix and perform teacher forcing on the concatenated sequence. Concretely, the decoder input is formed as $[\text{[BOS]};\mathbf{G}_{\text{prefix}};\mathbf{z}_{\text{pay}}]$, 
where $\mathbf{z}_{\text{pay}}=\phi(i_{\text{target}})$ denotes the token sequence of the target conversion item.

\begin{table*}[t]
\centering
\caption{Performance comparison on Tmall and Industry datasets. Best results are in \textbf{bold} and second-best are \underline{underlined}. "\textbf{Improv.}" shows the relative improvement (\%) over the second-best method.}
\label{tab:main}
\setlength{\tabcolsep}{10pt}
\resizebox{0.85\linewidth}{!}{
\begin{tabular}{ccccccccc}

\toprule
    \multirow{2}{*}{\textbf{Method}} & \multicolumn{4}{c}{\textbf{Industrial  Dataset}} & \multicolumn{4}{c}{\textbf{Tmall Dataset}} \\
\cmidrule(lr){2-5} \cmidrule(lr){6-9} 
     & Recall@5 & Recall@10 & NDCG@5 & NDCG@10 &  Recall@5 & Recall@10 & NDCG@5 & NDCG@10   \\ 
    \midrule \midrule
    
SASRec  & 18.67\% & 21.25\% & 14.49\% & 16.22\% & 39.36\% & 44.79\% & 30.66\% & 32.72\% \\
BERT4Rec& 18.32\% & 21.68\% & 14.41\% & 16.34\% & 40.47\% & 45.18\% & 31.39\% & 33.25\% \\
PBAT    & 19.35\% & 22.85\% & 14.96\% & 16.89\% & 40.21\% & 45.58\% & 31.16\% & 33.47\% \\
HSTU    & 20.24\% & 24.62\% & 15.57\% & 17.35\% & 40.63\% & 45.67\% & 31.60\% & 33.79\% \\
TIGER   & 21.40\% & 25.23\% & 16.28\% & 17.56\% & 42.61\% & 45.92\% & 32.96\% & 35.47\% \\
MBGen   & 23.39\% & 26.75\% & 18.23\% & \underline{19.35\%} & 44.10\% & \underline{47.57\%} & 35.14\% & \underline{37.94\%} \\
GEAR    & \underline{23.56\%} & \underline{26.87\%} & \underline{18.29\%} & 19.25\% & \underline{44.43\%} & 47.36\% & \underline{35.24\%} & 37.10\% \\
\midrule
\textbf{RCLRec (Ours)} & \textbf{26.21\%} & \textbf{29.62\%} & \textbf{20.80\%} & \textbf{21.91\%} & \textbf{49.78\%} & \textbf{52.37\%} & \textbf{40.36\%} & \textbf{42.48\%} \\
\textbf{Improv}.   & +11.25\% & +10.23\% & +13.72\% & +13.23\% & +12.04\%  & +10.09\% & +14.53\% & +11.97\% \\
\bottomrule
\end{tabular}
}
 \vspace{-10pt}
\end{table*}

\subsubsection{End-to-end curriculum optimization mechanism.} 
The above reverse curriculum prefix selection involves a discrete top-$k$ operation, which is non-differentiable. To enable end-to-end optimization, we introduce a curriculum optimization mechanism based on Index Backpropagation Quantization (IBQ) \cite{shi2025scalable}.


In the forward pass, we take the top-$k$ indices $\mathcal{K}$ from the relevance distribution $\mathbf{p}$ and build a hard $k$-hot mask $\mathbf{m}^{\text{hard}}\in\{0,1\}^T$ by setting $m^{\text{hard}}_t=\mathbb{I}(t\in\mathcal{K})$ for $t=1,\ldots,T$, where $\mathbb{I}(\cdot)$ is the indicator function. This mask specifies which historical positions are selected to construct the curriculum prefix $\mathbf{G}_{\text{prefix}}$ for the decoder.

In the backward pass, we replace the non-differentiable hard mask with a straight-through surrogate mask, enabling gradients to flow through index selection. Specifically, we construct a continuous selection vector $\mathbf{m} = \mathbf{m}^{\text{hard}} - \text{sg}[\mathbf{p}] + \mathbf{p}$,
where $\text{sg}[\cdot]$ is the stop-gradient operator. During the forward pass, $\mathbf{m}$ is identical to the hard mask $\mathbf{m}^{\text{hard}}$ and yields the same curriculum prefix. 

During the backward pass, gradients are propagated through $\mathbf{p}$ instead of the non-differentiable top-$k$ operator, effectively enabling end-to-end optimization of the selection distribution.
The resulting curriculum-augmented SFT loss is
\begin{small}
\begin{equation}
\mathcal{L}_{\text{SFT}}
= -\sum_{\ell=1}^{|\mathbf{G}_{\text{prefix}}|+L}
\log p_\theta\!\left(
\mathbf{y}^{\ell}
\mid
S_u,\,
\mathbf{e}_{\text{pay}},\,
\mathbf{y}^{<\ell}
\right),
\quad
\mathbf{y}=[\mathbf{G}_{\text{prefix}};\mathbf{z}_{\text{pay}}],
\label{eq:sft_curr}
\end{equation}
\end{small}
where $\mathbf{y}$ is the concatenated decoder-side sequence consisting of the curriculum prefix tokens $\mathbf{G}_{\text{prefix}}$ and the conversion target tokens $\mathbf{z}_{\text{pay}}$.
Thus, during SFT we apply teacher forcing on $\mathbf{y}$ and optimize the autoregressive likelihood of all decoder tokens.

\subsection{Curriculum Quality-Aware Loss}
\label{sec:loss}
Although RCPM injects selected intermediate curricula into the decoder, they may not always improve conversion modeling. Training the model with a curriculum prefix may increase its predictive confidence \cite{niculescu2005predicting}, but it does not guarantee improved conversion prediction performance over a no-curriculum baseline. We therefore define curriculum quality as the likelihood gain on conversion tokens brought by the selected prefix.

We first pre-train a standard encoder–decoder GR model (Eq.~(\ref{eq:gr})) to obtain parameters $\theta_0$. In the SFT stage, we initialize the curriculum-prefixed model from $\theta_0$ and optimize its parameters $\theta$ using Eq.~(\ref{eq:sft_curr}), while keeping $\theta_0$ as a fixed no-curriculum baseline when computing the quality-aware loss. 
We define the average conversion-token NLL with and without curriculum prefix as:
\begin{equation}
\begin{aligned}
\mathcal{L}^{\text{curr}}_{\text{pay}}(\theta)
&=
-\frac{1}{L}\sum_{\ell=1}^{L}
\log p_{\theta}\!\big(z_{\text{pay}}^{\ell} \mid S_u,\mathbf{e}_{\text{pay}},\mathbf{G}_{\text{prefix}},\mathbf{z}_{\text{pay}}^{<\ell}\big),\\
\mathcal{L}^{\text{base}}_{\text{pay}}(\theta_0)
&=
-\frac{1}{L}\sum_{\ell=1}^{L}
\log p_{\theta_0}\!\big(z_{\text{pay}}^{\ell} \mid S_u,\mathbf{e}_{\text{pay}},\mathbf{z}_{\text{pay}}^{<\ell}\big).
\end{aligned}
\end{equation}

A high-quality curriculum should reduce the conversion-token NLL compared with the baseline. We therefore define the curriculum quality-aware loss as a hinge constraint on the NLL reduction:
\begin{equation}
\mathcal{L}_{\text{qual}}
=
\max\Big(0,\, m-\big(\mathcal{L}^{\text{base}}_{\text{pay}}-\mathcal{L}^{\text{curr}}_{\text{pay}}\big)\Big),
\label{eq:l_qual}
\end{equation}
where $m\ge 0$ is a small margin. We jointly optimize this regularizer with the SFT objective in Eq.~(\ref{eq:sft_curr}), and the overall SFT loss is
\begin{equation}
\mathcal{L}_{\text{total}}
= \mathcal{L}_{\text{SFT}} + \lambda_{\text{qual}} \,\mathcal{L}_{\text{qual}},
\label{eq:l_total}
\end{equation}
where $\lambda_{\text{qual}}\ge 0$ controls the constraint strength. By explicitly maximizing conversion likelihood improvement over the fixed no-curriculum baseline, this loss steers RCPM toward selecting curricula that genuinely benefit conversion prediction.

\section{Experiments}
\subsection{Experimental Setup}

\textbf{Datasets.} We evaluate RCLRec on \textbf{two datasets}: (i) an in-house industrial advertising dataset from an Asian e-commerce platform with three behaviors and over 1B interactions, where conversions account for \textbf{1.23\%} of interactions, 
and (ii) the public Tmall dataset~\cite{zhong2015stock} with four behaviors using common preprocessing and user sampling settings, where conversions account for \textbf{0.95\%} of interactions. 
\textbf{Training \& evaluation metrics.}
We follow a pre-training and SFT pipeline.
We first pre-train on mixed multi-behavior data and then fine-tune only on conversion samples, using the same backbone and training settings. 
For offline evaluation, we report Recall@5/10 and NDCG@5/10 on the conversion target set \cite{esans, zheng2022multi}.

\textbf{Baselines \& implementation.}
We compare with \textbf{seven representative methods}: SASRec~\cite{sasrec}, BERT4Rec~\cite{bert4rec}, PBAT~\cite{su2023personalized}, HSTU~\cite{zhai2024actions}, TIGER~\cite{tiger}, MBGen~\cite{mbgen}, and GEAR~\cite{gear}. All baselines are implemented according to their original designs and trained on the same dataset partition for a fair comparison.
For RCLRec, each item is mapped to $L=4$ semantic tokens. We use a 4-layer Transformer encoder and a 2-layer decoder with hidden size $d=256$ and 8 attention heads. We set $k=4$, $\tau=0.5$, and $\lambda_{\text{qual}}=0.1$.

\begin{figure}[t]
  \captionsetup{skip=1pt}
  \centering
  \includegraphics[width=0.9\linewidth]{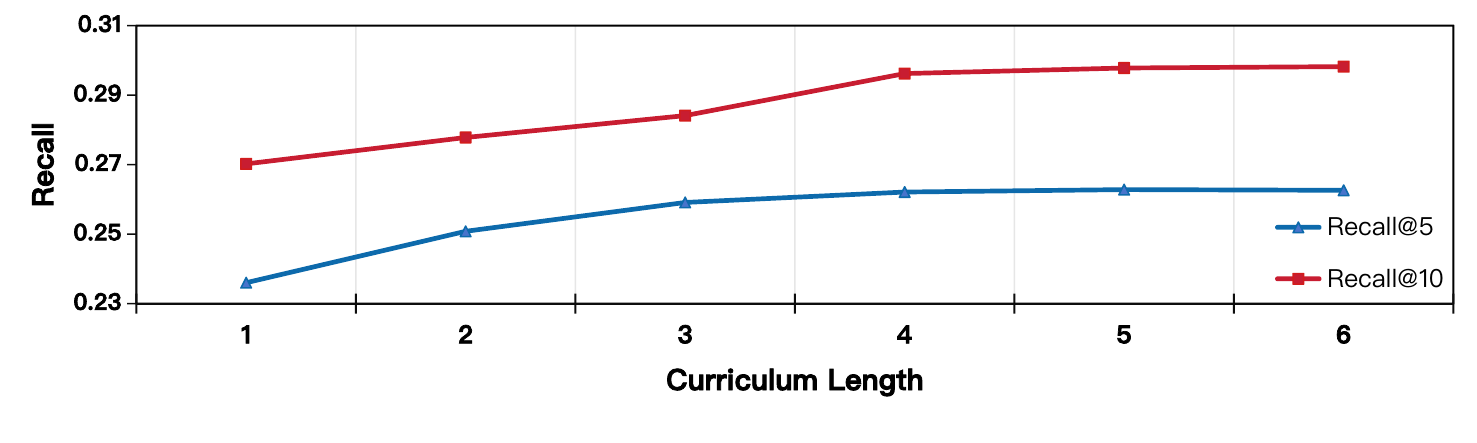} 
  \caption{Scalability and sensitivity of curriculum length $k$.}
  \vspace{-22pt}
  \captionsetup{skip=-2pt}
  \label{fig:topk}
\end{figure}
\subsection{Experimental Results}
\textbf{Overall performance.}
Table~\ref{tab:main} reports the overall performance on both datasets.
We observe that:
(1) RCLRec achieves the best Recall@5/10 and NDCG@5/10 on the conversion target set, outperforming all baselines. 
(2) Traditional sequential models (\textit{e.g.}, SASRec, BERT4Rec) lag behind, indicating limited ability to exploit conversion-relevant signals under sparse supervision.
(3) Compared with multi-behavior GR baselines (\textit{e.g.}, MBGen, GEAR), RCLRec provides more consistent gains by explicitly constructing conversion-related prefixes and guiding their effectiveness during SFT. 

\textbf{Scalability and sensitivity analysis.}
We vary the curriculum length $k\in\{1,\dots,6\}$  on the industrial dataset and report Recall@5/10, as shown in Figure~\ref{fig:topk}. 
A moderate curriculum size (around $k=4$) provides a significant improvement over shorter prefixes, while further increasing $k$ brings only marginal gains, suggesting that a small number of carefully selected interactions already capture most of the conversion-relevant context and that longer prefixes mainly introduce noise.
\begin{table}[htbp]
  \vspace{-8pt}
\centering
\caption{Ablation results on the industrial dataset. }
\label{tab:ablation}
\setlength{\tabcolsep}{2pt}
\small
\resizebox{0.85\linewidth}{!}{
\begin{tabular}{lcccc}
\toprule
\textbf{Variant} & \textbf{Recall@5} & \textbf{Recall@10} & \textbf{NDCG@5} & \textbf{NDCG@10}\\
\midrule
\textbf{RCLRec} & \textbf{26.21\%} & \textbf{29.62\%} & \textbf{20.80\%} & \textbf{21.91\%} \\
w/o RCPM  & 23.61\% & 26.78\% & 18.83\% & 19.89\% \\
Recent-$k$ curriculum & 24.57\% & 27.91\% & 19.37\%  & 20.49\%\\
w/o Quality-Aware Loss  & 25.55\% & 28.73\% & 20.15\%  & 21.13\%\\
\bottomrule
\end{tabular}
}
  \vspace{-10pt}
\end{table}

\textbf{Ablation study.}
Table~\ref{tab:ablation} summarizes the effect of each component.
Removing the RCPM (w/o RCPM) causes a large performance drop, indicating that explicitly constructing conversion-oriented curricula is crucial. Using recent curricula (Recent-$k$), which simply takes the most recent $k$ interactions, underperforms across all metrics, showing that the gain comes from conversion-oriented selection rather than recency.
Removing the curriculum quality-aware loss (w/o Quality-Aware Loss) also harms performance, suggesting that it improves curriculum usefulness and target relevance.

\textbf{Online results.}
We further evaluated RCLRec in an online A/B test on an industrial e-commerce advertising platform from Jan.\ 20 to Jan.\ 27, 2026. We used TIGER~\cite{tiger} as the baseline, with each bucket receiving 25\% of randomly sampled users. RCLRec improved \textbf{advertising revenue} by \textbf{2.09\%} and \textbf{orders} by \textbf{1.86\%}, demonstrating its effectiveness in production.

\section{Conclusion}
This paper presents \textbf{RCLRec}, a reverse curriculum learning–based generative recommendation (GR) framework for sparse conversion supervision. RCLRec introduces a Reverse Curriculum Prefix Module (RCPM), which traces backward from each conversion target, selects a short curriculum of key intermediate behaviors from the history, and feeds their semantic tokens to the decoder as a prefix together with the target tokens under a joint generation objective. This design transforms key intermediate behaviors into instance-specific supervision, alleviating conversion sparsity and guiding the model toward the user's critical decision process around conversions. We further propose a curriculum quality–aware loss that encourages the selected subsequences to be informative for conversion prediction. Extensive experiments on offline and online datasets show that RCLRec achieves superior performance and offers valuable insights for GR with sparse targets. In future work, we plan to extend the RCLRec framework to other target behaviors and richer curriculum designs.

\bibliographystyle{ACM-Reference-Format}
\bibliography{reference}
\end{document}